\documentstyle[11pt]{article}

\let\a=\alpha \let\b=\beta   
    
 \let\m=\mu \let\n=\nu

\def\nn{\nonumber} \def\bd{\begin{document}} \def\ed{\end{document}}
\def\ds{\documentstyle} \let\fr=\frac \let\bl=\bigl \let\br=\bigr
\let\Br=\Bigr \let\Bl=\Bigl 
\let\bm=\bibitem
\let\na=\nabla
\let\pa=\partial \let\ov=\overline 
\newcommand{\be}{\begin{equation}} 
\newcommand{\ee}{\end{equation}} 
\def\ba{\begin{array}}
\def\ea{\end{array}}
\def\ft#1#2{{\textstyle{{\scriptstyle #1}\over {\scriptstyle #2}}}}
\def\fft#1#2{{#1 \over #2}}
\def\del{\partial}
\def\vp{\varphi}
\def\sst#1{{\scriptscriptstyle #1}}
\def\oneone{\rlap 1\mkern4mu{\rm l}}
\def\td{\tilde}
\def\wtd{\widetilde}
\def\ie{\rm i.e.\ }
\def\dalemb#1#2{{\vbox{\hrule height .#2pt
        \hbox{\vrule width.#2pt height#1pt \kern#1pt
                \vrule width.#2pt}
        \hrule height.#2pt}}}
\def\square{\mathord{\dalemb{6.8}{7}\hbox{\hskip1pt}}}
\newcommand{\ho}[1]{$\, ^{#1}$}
\newcommand{\hoch}[1]{$\, ^{#1}$}
\newcommand{\bea}{\begin{eqnarray}} 
\newcommand{\eea}{\end{eqnarray}} 
\newcommand{\ra}{\rightarrow}
\newcommand{\lra}{\longrightarrow}
\newcommand{\Lra}{\Leftrightarrow}
\newcommand{\ap}{\alpha^\prime}
\newcommand{\bp}{\tilde \beta^\prime}
\newcommand{\tr}{{\rm tr} }
\newcommand{\Tr}{{\rm Tr} } 
\def\0{{\sst{(0)}}}
\def\1{{\sst{(1)}}}
\def\2{{\sst{(2)}}}
\def\3{{\sst{(3)}}}
\def\4{{\sst{(4)}}}
\def\5{{\sst{(5)}}}
\def\6{{\sst{(6)}}}
\def\7{{\sst{(7)}}}
\def\8{{\sst{(8)}}}
\def\n{{\sst{(n)}}}
\def\cA{{{\cal A}}}
\def\cF{{{\cal F}}}
\def\tV{\widetilde V}
\def\tW{\widetilde W}
\def\tH{\widetilde H}
\def\tE{\widetilde E}
\def\tF{\widetilde F}
\def\tA{\widetilde A}
\def\im{{{\rm i}}}
\def\tY{{{\wtd Y}}}
\def\ep{{\epsilon}}
\def\vep{{\varepsilon}}
\def\R{\rlap{\rm I}\mkern3mu{\rm R}}
\def\bD{{{\bar D}}}

\newcommand{\NP}{Nucl. Phys. }
\newcommand{\tamphys}{\it Center for Theoretical Physics,
Texas A\&M University, College Station, TX 77843}
\newcommand{\upenn}{\it Dept. of Phys. and Astro.,
University of Pennsylvania,
Philadelphia, PA 19104}

\newcommand{\auth}{I.Y. Park\hoch{1}, C.N. Pope\hoch{1} 
                   and A. Sadrzadeh\hoch{1}}

\begin{document}
\begin{flushright}
\hfill{CTP TAMU-30/01 \\
October 2001}\\
\hfill{\bf hep-th/0110238}\\
\end{flushright}

\vspace{10pt}

\begin{center}
{\large {\bf AdS and dS Braneworld Kaluza-Klein Reduction}}

\vspace{20pt}

\auth

\vspace{10pt}
{\hoch{\ddagger}\tamphys}

\vspace{30pt}

\underline{ABSTRACT}
\end{center}

    We obtain new results for consistent braneworld Kaluza-Klein
reductions, showing how we can derive four-dimensional $N=2$ gauged
supergravity ``localised on the AdS$_4$ brane'' as an exact embedding
in five-dimensional $N=4$ gauged supergravity.  Similarly, we obtain
five-dimensional $N=2$ gauged supergravity localised on an AdS$_5$
brane as a consistent Kaluza-Klein reduction from six-dimensional
$N=4$ gauged supergravity.  These embeddings can be lifted to type IIB
and massive type IIA supergravity respectively.  The new AdS
braneworld Kaluza-Klein reductions are generalisations of earlier
results on braneworld reductions to ungauged supergravities.  The
lower-dimensional cosmological constant in our AdS braneworld
reductions is an adjustable parameter, and so it can be chosen to be
small enough to be phenomenologically realistic, even if the
higher-dimensional one is of Planck scale.  We also discuss analytic
continuations to give a de Sitter gauged supergravity in four
dimensions as a braneworld Kaluza-Klein reduction.  We find that there
are two distinct routes that lead to the same four-dimensional theory.
In one, we start from a five-dimensional de Sitter supergravity, which
itself arises from a Kaluza-Klein reduction of type IIB$^*$
supergravity on the hyperbolic 5-sphere.  In the other, we start from
AdS gauged supergravity in five dimensions, with an analytic
continuation of the two 2-form potentials, and embed the
four-dimensional de Sitter supergravity in that.  The five-dimensional
theory itself comes from an $O(4,3)/O(3,2)$ reduction of Hull's 
type IIB$_{7+3}$ supergravity in ten dimensions.

{\vfill\leftline{}\vfill
\vskip 10pt 
{\footnotesize
        \hoch{1}        Research supported in part by DOE
grant DOE-FG03-95ER40917 \vskip -12pt}  \vskip  14pt
}

\pagebreak
\setcounter{page}{1}

\section{Introduction}

   An intriguing proposal that has attracted much attention in recent times
is the suggestion by Randall and Sundrum that four-dimensional physics 
can effectively arise from a five-dimensional theory that admits  
anti-de Sitter spacetime, but not Minkowski spacetime, as a vacuum 
\cite{ransun2}.  In the ground state four-dimensional Minkowski
spacetime is embedded as a ``3-brane'' in AdS$_5$ via the Poincar\'e 
or horospherical description, with
\be
d\hat s_5^2 = dz^2 + e^{-2k\, |z|}\, dx^\mu\, dx^\nu\, \eta_{\mu\nu}\,,  
\ee
where the AdS$_5$ metric $d\hat s_5^2$ satisfies $\hat R_{MN}= -4k^2\, 
\hat g_{MN}$ in the bulk.  The effective four-dimensional gravity can 
be investigated at the linearised level by replacing the Minkowski
metric $\eta_{\mu\nu}$ by $\eta_{\mu\nu} + h_{\mu\nu}$, and studying 
the equation governing small fluctuations.  This leads to the conclusion
that gravity is localised on the brane.

    One can consider the situation beyond the linearised level, by
introducing a general four-dimensional metric $ds_4^2$, and writing 
\be
d\hat s_5^2 = dz^2+ e^{-2k\, |z|}\, ds_4^2\,.\label{54start}
\ee
One can easily show that in the bulk the Ricci tensor of $d\hat s_5^2$ 
satisfies the Einstein equation $\hat R_{MN}=-4k^2\, \hat g_{MN}$ if the 
Ricci tensor of the four-dimensional metric satisfies $R_{\mu\nu}=0$
(see, for example, \cite{breper,hawcha,bob1,chasab1,chasab2}).
This approach was used, for example, in \cite{hawcha} to study the 
global 5-dimensional geometry resulting from having a four-dimensional 
Schwarzschild black hole ``on the brane.''

   Equation (\ref{54start}) has very much the structure of a
Kaluza-Klein reduction ansatz, and indeed one can view it as giving a
consistent embedding of $D=4$ pure Einstein theory with equation of
motion $R_{\mu\nu}=0$ in $D=5$ Einstein theory with a negative
cosmological constant, with equation of motion $\hat R_{MN}= -4k^2\,
\hat g_{MN}$.  More precisely, the embedding is an exact one if the
modulus sign is omitted in (\ref{54start}), giving
\be
d\hat s_5^2 = dz^2+ e^{-2k\, z}\, ds_4^2\,.\label{flatkk}
\ee
This now satisfies the five-dimensional equation not only in the bulk, but
also at $z=0$ itself, whereas previously (\ref{54start}) would have given 
a delta-function contribution to $\hat R_{MN}$, owing to the discontinuity
in the gradient of the metric there.

   This Kaluza-Klein theme was developed in \cite{bob1}, where it was shown 
that the above embedding could be extended to a fully consistent 
Kaluza-Klein reduction of $N=4$ gauged supergravity in $D=5$ to give 
$N=2$ ungauged supergravity in $D=4$.  The photon of the four-dimensional 
supergravity theory arises in a somewhat unusual way; rather than coming from
the $\hat g_{\mu 5}$ components of the five-dimensional metric, it 
comes instead from the reduction of the two 2-form gauge potentials 
$\hat A_\2^\a$ of the
five-dimensional $N=4$ gauged theory, with the ansatz \cite{bob1}
\be
\hat A_\2^1 = \ft1{\sqrt2}\, e^{-k\, z}\, F_\2\,,\qquad
\hat A_\2^2 = -\ft1{\sqrt2}\, e^{-k\, z}\, {*F_\2}\,,
\ee
where $*$ denotes the Hodge dual in four dimensions.  The equations of motion
of five-dimensional $N=4$ gauged supergravity then imply and are implied
by the equations of motion of four-dimensional $N=2$ ungauged supergravity.

   This consistent ``braneworld Kaluza-Klein reduction'' was also
generalised in \cite{bob1} to give five-dimensional $N=2$ ungauged 
supergravity from six-dimensional $N=4$ gauged supergravity, and to
give six-dimensional $N=1$ chiral ungauged supergravity from 
seven-dimensional $N=2$ gauged supergravity.  Further generalisations, 
including four-dimensional ungauged $N=4$ supergravity from five-dimensional
gauged $N=8$, and ungauged $N=2$ in six dimensions from gauged $N=4$ in seven,
were obtained in \cite{bob2}.  Some aspects of the fermion
reduction in the four-dimensional $N=2$ case were studied in \cite{dulisa}.

   Two universal features in all the braneworld Kaluza-Klein reductions
derived in \cite{bob1,bob2} were the halving of supersymmetry in the
reduction, as befits a theory localised on a brane, and also the fact
that a gauged supergravity in the higher dimension was reduced to give
an ungauged supergravity in the lower dimension.  This latter feature
was effectively built in from the outset, in the metric reduction
ansatz (\ref{flatkk}), which is based on the Poincar\'e or horospherical
embedding of (Minkowski)$_D$ spacetime in AdS$_{D+1}$. 

    Since one can also give a metric prescription for the embedding of
AdS$_D$ in AdS$_{D+1}$, this naturally raises the question of whether
one can generalise the constructions in \cite{bob1,bob2}, to obtain
braneworld Kaluza-Klein reductions that yield {\it gauged}
supergravities in the lower dimension from gauged supergravities in
the higher dimension.  This topic, which would be the Kaluza-Klein
counterpart of the viewpoint taken in \cite{karran}, forms the subject
of investigation in the present paper.

   The embedding of the metric $ds_D^2$ of AdS$_D$ in the metric 
$d\hat s_{D+1}^2$ of AdS$_{D+1}$ proceeds as follows:
\be
d\hat s_{D+1}^2 = dz^2 + \cosh^2 (k\, z)\, ds_D^2\,.\label{adsdd1}
\ee
A simple calculation shows that if $ds_D^2$ is Einstein, with
Ricci tensor given by
\be
R_{\mu\nu} = - (D-1)\, k^2\, g_{\mu\nu}\,,\label{ddricci}
\ee
then $d\hat s_{D+1}^2$ will be Einstein too, with Ricci tensor given by
\be
\hat R_{MN} = -D\, k^2\, \hat g_{MN}\,.\label{dd1ricci}
\ee

   It should be emphasised that although it might appear from 
(\ref{ddricci}) and (\ref{dd1ricci}) that there is a close correlation 
between the scale-size of the cosmological constants in the higher and the 
lower dimension, this is in fact not the case.  This can be seen by making
the constant coordinate shift $z =\td z- c$, together
with the constant scaling $ds_D^2=4 e^{-2k\, c}\, d\td s_D^2$.  If we now let
$\lambda=e^{-2k\, c}$, and drop the tildes, we have the metric 
reduction ansatz 
\be
d\hat s_{D+1}^2 = dz^2 + (\lambda\, e^{k\, z} + e^{-k\, z})^2\, ds_D^2\,,
\label{ads2}
\ee
for which we find that the $(D+1)$-dimensional and $D$-dimensional 
Ricci tensors are given by
\be
\hat R_{MN}= -D\, k^2\, \hat g_{MN}\,,\qquad
R_{\mu\nu} = -4\lambda\, (D-1)\, k^2\, g_{\mu\nu}\,.
\ee
Thus by choosing $\lambda$ appropriately, one can make the
cosmological constant of the lower-dimensional anti-de Sitter
spacetime as small as desired, no matter how large the cosmological
constant in the higher dimension.  For example, we could arrange to
have a four-dimensional cosmological constant that is compatible with
experimental observation, even with a Planck-scale five-dimensional
cosmological constant.  (Of course an extreme example is to take
$\lambda=0$, in which case the AdS embedding (\ref{ads2}) reduces to
the Poincar\'e embedding (\ref{flatkk}) of Minkowski spacetime in the
higher-dimensional AdS.)  In everything that follows in the rest of
this paper we shall similarly be able to tune the cosmological
constant (or, equivalently, the supergravity gauge coupling constant)
to any desired value in the lower dimension, regardless of the size of
the higher-dimensional cosmological constant.\footnote{We are grateful
to Hong L\"u for emphasising this point to us.}

   The embedding (\ref{adsdd1}), with $ds_D^2$ taken to be a general
$D$-dimensional metric, will form the metric reduction ansatz in our
braneworld Kaluza-Klein reductions.  We shall focus on two examples.
In the first, we shall show that four-dimensional gauged $N=2$
supergravity can be derived as a consistent braneworld reduction of
five-dimensional gauged $N=4$ supergravity.  We shall also show that
five-dimensional $N=2$ gauged supergravity can be obtained as a 
consistent braneworld reduction of six-dimensional $N=4$ gauged 
supergravity.  

   In each of our examples, the gauged supergravity that forms the
higher-dimensional starting point can itself be obtained as a
consistent Kaluza-Klein sphere reduction, from type IIB on $S^5$ in
the case of the five-dimensional theory \cite{d5gauge}, and from
massive IIA on a local $S^4$ in the case of the six-dimensional theory
\cite{d6gauge}.  By combining these sphere reductions with the
subsequent braneworld reductions, we can thereby obtain unusual
embeddings of the final four-dimensional and five-dimensional gauged
supergravities in type IIB and massive type IIA respectively.

   There has also been recent discussion of de Sitter spacetime, and
de Sitter braneworlds, in the literature \cite{malnun}-\cite{hull2}.
We show how, by a process of analytic continuation, we can obtain from
our first example an embedding of $N=2$ gauged de Sitter supergravity
in four dimensions as a braneworld reduction of $N=4$ gauged de Sitter
supergravity in $D=5$.  In fact in general one has the embedding
\be
d\hat s_{D+1}^2 = dz^2 + \cos^2 (k\, z)\, ds_D^2\,.\label{dsdd1}
\ee
of the $D$-dimensional de Sitter metric in $(D+1)$-dimensional de
Sitter spacetime, and this provides the starting-point for the
Kaluza-Klein de Sitter braneworld reduction.  The five-dimensional
starting point for a braneworld reduction to four-dimensional de
Sitter supergravity itself arises as a consistent Kaluza-Klein
reduction on the hyperbolic space $H^5$ of the type IIB$^*$ theory,
which was introduced by Hull \cite{hull1}. The type IIB$^*$ theory
arises by performing a T-duality transformation of type IIA on a
timelike circle, and is characterised by the fact that all the
Ramond-Ramond fields have kinetic terms of the ``wrong''
sign.\footnote{The usual IIA/IIB T-duality breaks down if one
considers reductions on timelike circles \cite{clpsst}, and so one can
view the type IIA$^*$ and type IIB$^*$ theories as necessary sectors
in the full string picture, which {\it must} be included in order to
regain duality when timelike reductions are included \cite{hull1}.}
It has been argued that this notwithstanding, the theory is a valid
sector within the web of dualities, with the subtleties of string
theory serving to protect it from the usual problems of negative
kinetic energies in field theory \cite{hull1}.  Recently, de Sitter
gauged supergravities related to the type IIA$^*$ and type IIB$^*$
theories have been discussed \cite{hull2}; they evade the usual
``no-go'' theorems for de Sitter supersymmetry precisely because of
the reversed signs for some of the kinetic terms.  As in the 
braneworld reductions to AdS supergravities, we can adjust the 
lower-dimensional cosmological constant at will, by performing 
rescalings of the type discussed above equation (\ref{ads2}).

   Our procedure for obtaining the de Sitter braneworld reduction
consists of first noting that we can obtain the type IIB$^*$ theory by an
analytic continuation of the usual type IIB theory, in which all the
Ramond-Ramond fields $\Psi$ are transformed according to 
$\Psi\longrightarrow \im\, \Psi$.  We then implement this continuation
in the usual $S^5$ reduction to $D=5$, thereby obtaining the gauged de Sitter
supergravity, with the $S^5$ analytically continuing to become $H^5$.  
Then, the analytic continuation of our previous braneworld reduction gives
the embedding of gauged $N=2$ de Sitter supergravity on the brane in
four dimensions.  Of course the existence of de Sitter supergravities
depends also upon the fact that certain fields have the ``wrong'' signs
for their kinetic terms.  By contrast, we find that our $D=6$ to $D=5$
AdS braneworld reduction does not admit an analytic continuation to a
real reduction of the six-dimensional gauged de Sitter supergravity
that can be obtained from reduction of the type IIA$^*$ theory.

   We also show that the same four-dimensional de Sitter gauged supergravity
can be obtained as a braneworld Kaluza-Klein reduction of a quite 
different type.  This is based on the fact that $D$-dimensional de Sitter
spacetime can be embedded within $(D+1)$-dimensional anti-de Sitter
spacetime, with the metric given by
\be
d\hat s_{D+1}^2 = dz^2 + \sinh^2(k\, z)\, ds_D^2\,.\label{ads33}
\ee
This has $\hat R_{MN}= -D\, k^2\, \hat g_{MN}$ if $R_{\mu\nu}= (D-1)\,
k^2\, g_{\mu\nu}$, and again, with a shift and rescaling of the type
discussed above equation (\ref{ads2}), the lower-dimensional
cosmological constant can be made arbitrary.  Using an ansatz
involving (\ref{ads33}) for the metric reduction, we obtain a
consistent braneworld Kaluza-Klein reduction from an anti-de Sitter
gauged $N=4$ supergravity in five dimensions, with an analytic
continuation of the two 2-form potentials, to give de Sitter gauged
$N=2$ supergravity in four dimensions.  This AdS$_5$ gauged
supergravity can itself be obtained from the type IIB$_{7+3}$ 
ten-dimensional supergravity introduced by Hull \cite{hull3}, which
has spacetime signature $(7,3)$, by reducing on the $(3,2)$-signature
space $O(4,2)/O(3,2)$.  The associated string theory is obtained from
the usual type IIB theory by means of a sequence of T-duality 
transformations involving spacelike and timelike circles.

   In the conclusions of this paper, we make some remarks about possible
applications of our results, and about the relation between
conventional braneworld scenarios and the Kaluza-Klein viewpoint.

\section{Gauged $N=2$, $D=4$ supergravity from $N=4$, $D=5$}
\label{d4n2sugra}

   In this section, we show how gauged $N=2$ supergravity in four
dimensions arises as a consistent Kaluza-Klein braneworld reduction from
gauged $N=4$ supergravity in five dimensions.  This generalises previous
work in \cite{bob1}, where it was shown how the ungauged $N=2$ theory
could be obtained as a consistent Kaluza-Klein braneworld reduction
from the same gauged $N=4$ starting point in $D=5$.  Our results here 
can be viewed as an exact and fully non-linear extension of the discussion 
of supergravity fluctuations on an AdS$_4$ braneworld embedded in $D=5$.  This 
generalises the interpretation of the Kaluza-Klein reduction in 
\cite{bob1}, which was the analogous non-linear extension of supergravity
fluctuations around a flat Minkowski$_4$ braneworld.

\subsection{The bosonic fields}\label{bossec}

   Our starting point is the $SU(2)\times U(1)$ gauged $N=4$ supergravity
in five dimensions.  The bosonic sector of the
five-dimensional theory comprises the metric, a dilatonic scalar
$\phi$, the $SU(2)$ Yang-Mills potentials $\hat A^i_\1$, a $U(1)$ gauge
potential $\hat B_\1$, and two 2-form potentials $\hat A^\a_\2$ which transform
as a charged doublet under the $U(1)$.  The Lagrangian \cite{wnc}, 
expressed in the language of differential forms that we shall use here,
is given by \cite{d5gauge}
\bea
{\cal L}_5 &=& \hat R\, {\hat *\oneone} - \ft12{\hat *d\phi}\wedge d\phi
-\ft12 X^4\,  {\hat *\hat G_\2}\wedge \hat G_\2
-\ft12 X^{-2}\, ({\hat *\hat F^i_\2}\wedge \hat F^i_\2 
+ {\hat*\hat A^\a_\2}\wedge \hat A^\a_\2)\nn\\
& & +\fr1{2g} \epsilon_{\a\b}\, \hat A^\a_\2\wedge d\hat A^\b_\2 - \ft12
\hat A^\a_\2\wedge \hat A_\2^\a\wedge \hat B_\1 
- \ft12 \hat F^i_\2\wedge \hat F^i_\2\wedge \hat B_\1\nn\\
& &  + 4 g^2\, (X^2 + 2 X^{-1})\, {\hat *\oneone}\,,
\label{d5lag}
\eea
where $X=e^{-\ft1{\sqrt6}\, \phi}$, $\hat F_\2^i = d\hat A_\1^i + \ft1{\sqrt2}
g\, \ep^{ijk}\, \hat A_\1^j\wedge \hat A_\1^k$ and 
$\hat G_\2=d\hat B_\1$, and $\hat *$
denotes the five-dimensional Hodge dual. It is useful to
adopt a complex notation for the two 2-form potentials, by defining
\be
A_\2 \equiv A_\2^1 + \im\, A_\2^2\,.\label{complexa}
\ee

    Our bosonic Kaluza-Klein reduction ansatz involves setting the
fields $\phi$ and $B_\1$ to zero, together with two out of the three
components of the $SU(2)$ Yang-Mills fields $A_\1^i$.  We find that
the ansatz for the remaining bosonic fields, comprising the metric,
the two 2-form potentials, and the surviving Yang-Mills potential,
which we shall take to be $A_\1^1$, is
\bea
d\hat s_5^2 &=& dz^2 + \cosh^2(k\, z)\, ds_4^2\,,\nn\\
\hat A_\2^1 &=& -\ft1{\sqrt2}\, \cosh(k\, z)\, {*F_\2}\,,\qquad 
\hat A_\2^2 = \ft1{\sqrt2}\, \sinh(k\, z)\, F_\2 \,,\nn\\
\hat A_\1^1 &=& \ft1{\sqrt2}\, A_\1\,,\label{d5ansatz}
\eea
where $ds_4^2$ is the metric and $F_\2=dA_\1$ is the Maxwell field of the
four-dimensional $N=2$ supergravity, and $*$ denotes the Hodge dual in
the four-dimensional metric.

   We now substitute this ansatz into the five-dimensional equations
of motion that follow from (\ref{d5lag}), namely \cite{d5gauge,bob1}
\bea
d(X^{-1}\, {\hat * dX}) &=& \ft13 X^4\,  {\hat * \hat G_\2}\wedge \hat G_\2 -
\ft16 X^{-2} \, ({\hat * \hat F^i_\2}\wedge \hat F^i_\2 + 
{\hat * \bar{ \hat A_\2}}\wedge \hat A_\2)\nn\\
& & - \ft{4}{3}g^2\, (X^2 - X^{-1})\, {\hat * \oneone} ,\nn\\
d(X^4\, {\hat * \hat G_\2}) &=& - \ft12 \hat F^i_\2\wedge \hat F^i_\2 -
\ft12 {\bar {\hat A}}_\2\wedge \hat A_\2 ,\nn\\
d(X^{-2}\, {\hat * \hat F^i_\2}) &=& \sqrt{2}\, g \,
\epsilon^{ijk}\, X^{-2}\, {\hat * \hat F^j_\2}\wedge \hat A^k_\1
- \hat F^i_\2\wedge \hat G_\2,\nn\\
X^{2}\, {\hat * \hat F_\3} &=& - \im \, g\, \hat A_\2 \,,\nn\\
\hat R_{MN} &=& 3 X^{-2}\,  \pa_M X\, \pa_N X - \ft{4}{3}g^2\,(X^2 +
2  X^{-1})\, \hat g_{MN}\nn\\
& & + \ft12 X^4 \, (\hat G_M{}^P \hat G_{N P} -\ft16 \hat g_{MN} \,
\hat G_\2^2) + \ft12 X^{-2}\,
(\hat F^{i\ P}_M \hat F^{i}_{NP}  - \ft16 \hat g_{MN}\, (\hat F^i_\2)^2)\nn\\
& & + \ft12 X^{-2}\,  ({\bar{ \hat A}}_{(M}{}^P\,  \hat A_{N)P} - \ft16
\hat g_{MN}\, |\hat A_\2|^2)\,,
\label{d5eom}
\eea 
where
\be
\hat F_\3 =D \hat A_\2 \equiv d\hat A_\2  -\im\, g\, 
          \hat B_\1\wedge \hat A_\2\,.\label{gauged}
\ee
In order to do this, it is useful to record that the Ricci tensor 
$\hat R_{AB}$ for the five-dimensional metric $d\hat s_5^2$ is related to
the Ricci tensor $R_{ab}$ of the four-dimensional metric $ds_4^2$ by
\bea
\hat R_{ab} &=& {\rm sech}^2(k\, z)\, R_{ab} - 3 k^2\, \tanh^2(k\, z) \, 
                \eta_{ab} -k^2\, \eta_{ab}\,,\nn\\
\hat R_{55} &=& -4k^2\,,\qquad \hat R_{a5}=0\,,
\eea
where $A=(a,5)$, and we are using tangent-space indices.  The following 
{\it lemmata} are also helpful:
\bea
&&\bar{\hat A_\2}\wedge \hat A_\2 = -\ft12 F_\2\wedge F_\2\,,\qquad
 {\hat *\bar{\hat A_\2}}\wedge \hat A_\2 = -\ft12 {*F_\2}\wedge F_\2
 \wedge dz\,,\nn\\
&&{\hat * F_\2} = {*F_\2}\wedge dz\,,\qquad {\hat *(F_\2\wedge dz)} 
={* F_\2}\,.
\eea

  After substituting the braneworld reduction ansatz, we find that all
the five-dimensional equations of motion are satisfied, with an exact
cancellation of all dependence on the fifth coordinate $z$, if and only 
if the four-dimensional fields satisfy the bosonic equations of motion of 
four-dimensional gauged $N=2$ supergravity, namely
\bea
R_{\mu\nu} &=& \ft12 (F_{\mu\rho}\, F_\nu{}^\rho - 
         \ft14 F_\2^2 \, g_{\mu\nu}) - 3k^2\, g_{\mu\nu}\,,\nn\\
d{*F_\2} &=& 0\,,\label{d4eom}
\eea
with
\be
 k=g\,.
\ee
It should be emphasised that the cancellation of the $z$ dependence in 
the five-dimensional equations of motion is quite non-trivial, and it depends 
crucially on the details of the ansatz (\ref{d5ansatz}).  In particular, 
the presence of the four-dimensional $U(1)$ gauge field both in the
2-form potentials $\hat A_\2$ and in a $U(1)$ subgroup of the $SU(2)$ 
Yang-Mills fields $\hat A_\1^i$ in five dimensions is essential for the
matching of the $z$ dependence to work.  We should also emphasise
that, as discussed in the introduction, the cosmological constant in
the four-dimensional gauged theory can be adjusted at will, whilst
keeping the five-dimensional cosmological constant fixed, by 
shifting to a new coordinate $z=\td z-c$, together  with the rescaling
$ds_4^2=4 e^{-2k\, c}\, d\td s_4^2$ of the four dimensional metric,
and now, in addition, we rescale the four-dimensional gauge field so 
that $A_\1=2 e^{-k\, c}\, \td A_\1$.  We shall return to this point in
section \ref{d4ungauged}. 

\subsection{The fermionic fields}\label{d54fermion}

   Having obtained a consistent Kaluza-Klein reduction ansatz in the
bosonic sector, we now turn to the fermions.  In the five-dimensional
$N=4$ gauged theory there are spin-$\ft12$ fields $\hat \chi_p$, where
$p $ is a 4-dimensional $USp(4)$ index, and spin-$\ft32$ fields
$\hat\psi_{M\, p }$.  Our fermionic ansatz will involve setting
$\hat\chi_{\a}$ to zero, and so we must first verify that this is
compatible with the expected surviving $N=2$ supersymmetry in $D=4$.
Taking into account that certain of the bosonic fields have already
been set to zero in the reduction ansatz, the remaining non-vanishing
five-dimensional fields have the following contributions\footnote{In
common with the majority of recent work in supergravity, we shall
neglect quartic fermion terms in the Lagrangian, and their associated
consequences in the supersymmetry transformation rules.} in the
supersymmetry transformation rule for $\hat\chi_p$:
\be
\delta\, \hat\chi_p = -\ft1{4\sqrt6}\, \gamma^{MN}\, \Big(F_{MN}^i\, 
(\Gamma_i)_{pq}\, + A^\a_{MN}\, (\Gamma_{\a})_{pq}\Big)\, \hat\ep^q\,,
\ee
where $\gamma_M$ are the five-dimensional spacetime Dirac matrices,
and $(\Gamma_\a,\Gamma_i)$ are the five ``internal'' $USp(4)\sim SO(5)$
Dirac matrices.\footnote{The five-dimensional index range is spanned
by $\a$, the doublet index on the 2-form potentials $\hat A_\2^\a$,
and $i$, the Yang-Mills $SU(2)$ triplet index on $\hat A_\1^i$.  In
what follows, we shall take $\a=(1,2)$, and $i=(3,4,5)$.  Thus our
previous bosonic ansatz with $\hat A_\1^i$ non-zero will now be
translated, in this fermionic discussion, to having a non-vanishing
Yang-Mills term for the index value $i=3$.}
Substituting the bosonic ansatz into this, we find
that for $\delta\, \hat\chi_p$ to vanish we must have
\be
\gamma^{\mu\nu}\, \Big(F_{\mu\nu}\,\Gamma_3 -\cosh(k\, z)\, 
({*F})_{\mu\nu} \, \Gamma_1 + \sinh(k\, z)\, F_{\mu\nu}\, 
    \Gamma_2\Big)_{pq}\, \hat \ep^q=0\,.
\ee
Using $\gamma^{\mu\nu}\, ({*F})_{\mu\nu} = \im\, \gamma^{\mu\nu}\,
\gamma_5\, F_{\mu\nu}$, we therefore deduce that the five-dimensional
supersymmetry parameters $\hat\ep_p$ must satisfy
\be
\Big(\Gamma_3 -\im\, \cosh(k\, z)
\, \gamma_5\, \Gamma_1 + \sinh(k\, z)\,\Gamma_2\Big)\, \hat\ep=0\,,
\label{chicon}
\ee
where we have now suppressed the $USp(4)$ internal spinor index $p$.  
We shall return to this equation shortly.

    For the five-dimensional gravitino transformation rule, taking
into account that certain bosonic fields are set to zero in the
reduction ansatz, we shall have
\bea
\delta\, \hat \psi_{M\, p} &=& \hat\nabla_M\, \hat\ep_p  +\Big(
\fft{g}{\sqrt2}\, \hat A_\1^i\, \Gamma_{12 i} 
    -\fft{\im\, g}{2}\, \gamma_M \, \Gamma_{12} \nn\\
&&
\qquad\qquad -\ft{\im}{12\sqrt2}\, 
  [\gamma_M{}^{PQ} - 4\delta_M^P\, \gamma^Q]\,
(\hat F_{PQ}^i\, \Gamma_i+ \hat A^\a_{PQ}\,
    \Gamma_\a)\Big)_{pq}\, \hat\ep^q\,.\label{psitrans}
\eea
Our fermion reduction ansatz will involve setting $\hat\psi_{M\, p}$
to zero in the $M=5$ direction.  The resulting requirement 
$\delta\, \hat\psi_{5\, p}=0$ for supersymmetry therefore leads to
\be
\del_z\, \hat\ep + \fft{\im\, g}{2}\, \gamma_5\, \Gamma_{12}\, \hat\ep 
=0\,,\label{psi5con}
\ee
where we have again suppressed the $USp(4)$ index on $\hat\ep$
$\gamma_5$ is $-\gamma_z$.
Bearing in mind that $g=k$, we now find that (\ref{chicon}) and
(\ref{psi5con}) are solved by taking
\be
\hat\ep = \Big( \cosh(\ft12 k\, z) -\im\, \sinh(\ft12 k\, z)\, 
  \gamma_5\, \Gamma_{12}\Big)\,\ep\,,\label{epans}
\ee
where $\ep$ is the four-dimensional supersymmetry parameter,
which must satisfy the constraint
\be
\gamma_5\, \Gamma_{13}\, \ep= \im\, \ep\,.\label{epcon}
\ee
This constraint has the effect of halving the supersymmetry. This is
consistent with the fact that we are starting from the $N=4$ gauged
theory in $D=5$, and ending up with the $N=2$ gauged theory in $D=4$.

    By examining the components of $\delta\, \hat\psi_{M\, p}$ when
$M$ lies in the four-dimensional spacetime, we are now in a position
to deduce the correct reduction ansatz for the gravitino. We find that
it is
\be
\hat\psi_\mu = \Big(\cosh(\ft12 k\, z) - \im\, \gamma_5\, \Gamma_{12}\,
\sinh(\ft12 k\, z)\Big)\, \psi_\mu\,,\label{psians}
\ee
where $\psi_\mu$ denotes the four-dimensional gravitini, subject also
to the constraint
\be
\gamma_5\, \Gamma_{13}\, \psi_\mu= \im\, \psi_\mu\,.\label{psicon}
\ee
Substituting (\ref{epans}) and (\ref{psians}) into (\ref{psitrans}), we
find that the $z$ dependence matches on the two sides of the equation
and we consistently read off the four-dimensional gravitino
transformation rule
\be
\delta\, \psi_\m = D_\mu\, \ep - \ft{1}{2}\, k\, \gamma_\mu\, \ep 
    +\ft{1}{8}\, F^{\nu\rho}\, \gamma_\mu\, \gamma_{\nu\rho}\,
\Gamma_{123}\, \ep\,,
\ee
where
\be
D_\mu\, \ep = \nabla_\mu\, \ep +\ft{1}{2}\, k\, A_\mu\,
\Gamma_{123}\, \ep\,.\label{gaugecov}
\ee

\subsection{The $D=4$ cosmological constant and the 
ungauged limit}\label{d4ungauged}

   If we implement the rescalings discussed in the introduction and at
the end of section \ref{bossec}, namely
\be
z=\td z -c\,,\qquad g_{\mu\nu}= 4 e^{-2k\, c}\, \td g_{\mu\nu}
\,,\qquad A_\mu = 2 e^{-k\, c}\, \wtd A_\mu\,,\label{rescale}
\ee
then after defining $\lambda\equiv e^{-2k\, c}$, and dropping the
tildes on the rescaled four-dimensional fields, we see that the bosonic
reduction ansatz (\ref{d5ansatz}) becomes
\bea
d\hat s_5^2 &=& dz^2 + (\lambda\, e^{k\, z} +e^{-k\, z})^2
\, ds_4^2\,,\nn\\
\hat A_\2^1 &=& -\ft1{\sqrt2}\, (\lambda\, e^{k\, z} + 
e^{-k\, z})\, {*F_\2}\,,\qquad 
\hat A_\2^2= \ft{1}{\sqrt2}\, (\lambda\, e^{k\, z} -e^{-k\,
z})\, F_\2  \,,\nn\\
\hat A_\1^1 &=& \sqrt{2\lambda}\, A_\1\,,\label{d5ansatz0}
\eea
and the resulting four-dimensional equations of motion (\ref{d4eom}) 
become
\bea
R_{\mu\nu} &=& \ft12 (F_{\mu\rho}\, F_\nu{}^\rho - 
         \ft14 F_\2^2 \,g_{\mu\nu}) - 12\lambda\, k^2\, 
g_{\mu\nu}\,,\nn\\
d{*F_\2} &=& 0\,,\label{d4eom0}
\eea
An analogous rescaling can also be performed in the fermionic
reduction ansatz obtained in (\ref{d54fermion}).

   The above rescaling shows that by choosing the constant $\lambda$
appropriately, we can obtain any desired value of the cosmological
constant in the four-dimensional gauged supergravity, whilst keeping
the five-dimensional cosmological constant fixed. In particular, we
could choose $\lambda$ so that the four-dimensional cosmological
constant is compatible with observation, even if the five-dimensional
cosmological constant is of Planck scale.  Of course the standard
supergravity relation between the cosmological constant and the gauge
cooupling constant in $D=4$ will still hold.  This can be seen, for
example, from the gauge-covariant derivative (\ref{gaugecov}), which
now becomes
\be
D_\m\, \ep = \nabla_\mu\, \ep + k\, \sqrt{\lambda}\, A_\mu\,
\Gamma_{123}\, \ep\,.
\ee

   We may also note that we can recover the results in \cite{bob1} for
the brane-world reduction to ungauged $N=2$ supergravity,\footnote{The
four-dimensional field strength here is the Hodge dual of the one
arising in \cite{bob1}.  The choice of duality complexion was
immaterial in \cite{bob1}, since the 1-form potential $A_\1$ itself
did not appear in the reduction ansatz there.  In the AdS braneworld
reduction we are considering in this paper the choice becomes
important, because the ansatz (\ref{d5ansatz}) involves the explicit
appearance of $A_\1$ in the Yang-Mills sector, which was not excited
in the ungauged case.}, by taking $\lambda=0$.  Note in particular
that the cosmological constant in the Einstein equation scales to zero
in this limit, as expected for the ungauged limit.  A similar
rescaling procedure applied to the fermionic sector shows that one
also recovers the fermionic results discussed in \cite{dulisa}.

\subsection{Lifting to type IIB in $D=10$}\label{s5redn}

    The $SU(2)\times U(1)$ gauged $N=4$ supergravity in $D=5$ that
formed the starting point for our braneworld Kaluza-Klein reduction to
$D=4$ can itself be obtained in a Kaluza-Klein reduction from type IIB
supergravity.  It is believed, although it has never been proved, that
the maximal $SO(6)$-gauged $N=8$ theory in $D=5$ arises as an $S^5$
reduction from type IIB.  For our purposes, it suffices to work with
a truncated $S^5$ reduction in which only the fields of $N=4$
supergravity in $D=5$ are retained, and in the bosonic sector this was
constructed in explicit detail in \cite{d5gauge}.  Using these
results, we may therefore lift our braneworld Kaluza-Klein reduction
to $D=10$, giving a novel way of embedding four-dimensional $N=2$
gauged supergravity as a consistent Kaluza-Klein reduction from type
IIB supergravity.

   There is no
simple covariant Lagrangian for type IIB supergravity, on account of
the self-duality constraint for the 5-form.  However, one can write a
Lagrangian in which the 5-form is unconstrained, which must then be
accompanied by a self-duality condition which is imposed by hand at
the level of the equations of motion \cite{bergs}.  This type IIB
Lagrangian, in the notation we shall use here, is \cite{d5gauge}
\bea
{\cal L}^{IIB}_{10} &=& \hat R\, {\hat *\oneone} - \ft12
{\hat *d\hat{\phi}}\wedge d\hat{\phi} - \ft12
e^{2\hat{\phi}}\,
{\hat *d\hat{\chi}}\wedge d\hat{\chi} - \ft14 {\hat *\hat H_\5}
\wedge \hat H_\5 \nn\\
& & -\ft12 e^{-\hat \phi} \, {\hat * \hat{F}^{2}_\3}\wedge\hat{F}^{2}_\3 -
\ft12 e^{\hat\phi}\, {\hat *\hat{F}^{1}_\3}\wedge \hat{F}^{1}_\3 -
\ft12 \hat B_\4\wedge d\hat{A}_\2^{1}\wedge d\hat{A}_\2^{2}\ ,\label{d10lag}
\eea
where $\hat F^2_\3 = d\hat A^2_\2,\, \hat F^1_\3=d\hat
A^1_\2 - \hat{\chi}\, d\hat{A}^2_\2,\, \hat{H}_\5=d\hat{B}_\4 -
\ft12 \hat{A}_\2^1\wedge d\hat{A}_\2^2 + \ft12
\hat{A}_\2^2\wedge d\hat{A}_\2^1$, and we use hats to denote
ten-dimensional fields and the ten-dimensional Hodge dual $\hat*$.  
The bosonic equations follow from the Euler-Lagrange equations,
together with the self-duality constraint $\hat H_\5={\hat *H_\5}$.  
The bosonic reduction ansatz is then given by \cite{d5gauge}
\bea
d\hat s^2_{10} &=&
\Delta^{1/2}\, ds_5^2 +  g^{-2} \, X\,
\Delta^{1/2}\, d\xi^2 +
g^{-2}\Delta^{-1/2}\, X^2\, s^2\, \Big(d\tau - g\, B_\1\Big)^2 \nn\\
& &+ \ft14 g^{-2}\, \Delta^{-1/2}\, X^{-1}\,
c^2\, \sum_i (\sigma^i - \sqrt{2}g\, A_\1^i)^2\ ,\nn\\
\hat{G}_\5 &=& 2g\, U \,\varepsilon_5 - \frac{3 sc}{g} X^{-1}\,
{*dX}\wedge d\xi +
\frac{c^2}{8\sqrt{2}\, g^2} X^{-2}\, {*F^i_\2}\wedge h^j\wedge h^k
\, \varepsilon_{ijk} \nn\\
& &-\frac{sc}{2\sqrt{2}\,g^2} X^{-2}\, {*F^i_\2}\wedge h^i\wedge d\xi
- \frac{sc}{g^2} X^4\, {*G_\2}\wedge d\xi\wedge (d\tau - g B_\1),\nn\\
\hat{A}_\2 &\equiv& \hat A_\2^1 + \im\, \hat A_\2^2 =
 -\fr{s}{\sqrt{2}g}\, e^{-\im\,\tau}\, A_\2\,,\nn\\
\hat{\phi} &=& 0,\ \ \ \hat{\chi} = 0,
\label{d105ans}
\eea
where $\hat H_\5=\hat G_\5 + {\hat *\hat G_\5}$, $h^i\equiv \sigma^i -
\sqrt{2} g\, A_\1^i$, $\Delta\equiv X^{-2}\, s^2 + X\, c^2$, $U\equiv
X^2\, c^2 + X^{-1}\, s^2 + X^{-1}$, $\ep_5$ is the volume form in the
five-dimensional spacetime metric $ds_5^2$, and $c\equiv \cos\xi$,
$s\equiv \sin\xi$.  The $\sigma_i$ are the left-invariant 1-forms of
$SU(2)$, and the 5-sphere on which the reduction is performed is
described, if the Yang-Mills and scalar field are taken to vanish, by
the round metric $ds^2 = d\xi^2 + s^2\, d\tau^2 + \ft14
c^2\,\sigma_i^2$.  Thus $S^5$ is viewed as a foliation by $S^1\times
S^3$.

   Owing to a lack of a variety of suitable adornments for fields, in
the above we are using the ``hat'' to denote ten-dimensional
quantities, while the unhatted quantities are five-dimensional.  We
now substitute our previous braneworld ansatz (\ref{d5ansatz}) into
this (taking appropriate care over the change of r\^oles of hatted and
unhatted fields), thereby obtaining a reduction ansatz from $D=10$ to
$D=4$.  This is therefore given by
\bea
d\hat s^2_{10} &=&dz^2 + 
\cosh^2(k\, z)\,  ds_4^2 +  g^{-2} \, \Big(d\xi^2 + s^2\, d\tau^2 
+ \ft14 
c^2\, (\sigma_1^2 + \sigma_2^2  +(\sigma_3- g\, A_\1)^2)\Big)\ ,\nn\\
\hat{G}_\5 &=& 4g\, \cosh^4(k\, z)\, dz\wedge \varepsilon_4 +
\frac{c^2}{8g^2}\, {*F_\2}\wedge dz\wedge \sigma_1\wedge \sigma_2
 \nn\\
& &-\frac{sc}{4g^2} \, {*F_\2}\wedge dz\wedge 
       (\sigma_3-g\, A_\1)\wedge d\xi\nn\\
\hat{A}_\2 &\equiv& \hat A_\2^1 + \im\, \hat A_\2^2 =
 \fr{s}{2g}\, e^{-\im\,\tau}\, (-\im\, \sinh(k\, z)\, F_2 + \cosh(k\,
 z)\, {*F_\2})\,,\nn\\
\hat{\phi} &=& 0,\ \ \ \hat{\chi} = 0,
\label{d104ans}
\eea

\section{Gauged $N=2$, $D=5$ supergravity from $N=4$, $D=6$}

   In this section, we show how a similar consistent braneworld
Kaluza-Klein reduction of the $SU(2)$ gauged $N=4$ theory in six
dimensions is possible, yielding $N=2$ (minimal) gauged supergravity
in $D=5$.  Again, this generalises a braneworld reduction in
\cite{bob1}, where the ungauged $N=2$ theory in five dimensions
was obtained.  Since the ideas used in the reduction to $D=5$ are very
similar to those for the reduction to $D=4$ in the previous section,
we shall give a rather briefer presentation of our results here.

\subsection{The bosonic fields}

   The bosonic fields in the six-dimensional $SU(2)$ gauged theory
comprise the metric, a dilaton $\hat\phi$, a 2-form potential $\hat A_\2$, and
a 1-form potential $\hat B_\1$, together with the $SU(2)$ gauge potentials 
$\hat A_\1^i$.  The bosonic Lagrangian \cite{romans6},
converted to the language of differential forms, is \cite{d6gauge}
\bea
{\cal L}_6 &=& \hat R\, {{\hat*}\oneone} -
\ft12 {{\hat *}d\hat\phi}\wedge d\hat\phi
- g^2\Big(\ft29 X^{-6} -\ft83 X^{-2} -
2 X^2\Big)\,  {{\hat *}\oneone}\nn\\
&&-\ft12 X^4\, {{\hat *}\hat F_\3\wedge \hat F_\3} -\ft12
X^{-2}\, \Big( {{\hat *}\hat G_\2}\wedge \hat G_\2 + 
{{\hat *}\hat F_\2^i}\wedge
\hat F_\2^i \Big) \label{d6lag}\\
&& - \hat A_\2\wedge(\ft12
d\hat B_\1\wedge d\hat B_\1 +\ft13 g\, \hat A_\2\wedge d\hat B_\1 +\ft2{27}
g^2\, \hat A_\2\wedge \hat A_\2 +\ft12 \hat F_\2^i\wedge \hat F_\2^i)\,,\nn
\eea
where $X\equiv e^{-\hat \phi/(2\sqrt2)}$, $\hat F_\3=d\hat A_\2$,
$\hat G_\2= d\hat B_\1 +
\ft23g\, \hat A_\2$, $\hat F_\2^i = d\hat A_\1^i + \ft12 g\, 
\ep_{ijk} \hat A_\1^j\wedge
\hat A_\1^k$, and here ${\hat *}$ denotes the six-dimensional Hodge dual. 
The resulting bosonic equations of motion are given in \cite{d6gauge}.

   We find that the following bosonic ansatz yields a consistent
braneworld Kaluza-Klein reduction:
\bea
d\hat s_6^2 &=& dz^2 + \cosh^2(k\, z)\, ds_5^2\,,\nn\\
\hat A_\2 &=& -\ft1{\sqrt3}\, k^{-1}\, \sinh(k\,z)\, F_\2\,,\nn\\
\hat A_\1^3 &=& -\sqrt{\ft23}\, A_\1\,,\nn\\
\hat B_\1 &=&0\,,\quad \hat A_\1^1=0\,,\qquad \hat A_\1^2=0\,,\quad
\hat\phi=0\,,\label{d6ans}
\eea
where $F_\2=dA_\1$ is the graviphoton of the five-dimensional $N=2$
supergravity, and 
\be
k=\ft{\sqrt2}{3}\, g\,.
\ee
Note that although $\hat B_\1$ is set to zero, the field strength
$\hat G_\2$, defined above, is non-zero, and is given by
\be
\hat G_\2 = -\sqrt{\ft23}\, \sinh(k\, z)\, F_2\,.
\ee
It should also be noted that, as in the reduction from $D=5$ to $D=4$,
we find it necessary for the five-dimensional graviphoton to appear
not only in the ansatz for $\hat A_\2$, but also in 
one component of the $SU(2)$ Yang-Mills fields in $D=6$.  

    After substituting the ansatz (\ref{d6ans}) into the equations of
motion in \cite{d6gauge} that follow from (\ref{d6lag}), we find that
all the $z$ dependence matches in a consistent fashion, yielding the
following five-dimensional equations of motion:
\bea
R_{\mu\nu} &=& \ft12(F_{\mu\rho}\, F_\nu{}^\rho - \ft16 F_\2^2\,
g_{\mu\nu}) - 4 k^2\, g_{\mu\nu}\,,\nn\\
d{*F_\2} &=& \ft1{\sqrt3}\, F_\2\wedge F_\2\,.\label{d5sugra}
\eea
These are precisely the bosonic equations of motion of $N=2$ (\ie
minimal) gauged supergravity in five dimensions.  We may note that, as
in section \ref{d4ungauged}, we can make the analogous rescaling
(\ref{rescale}), so that the five-dimensional cosmological constant
becomes freely adjustable.  In particular, by setting $\lambda=0$ we
recover the braneworld Kaluza-Klein reduction to the ungauged
supergravity that was obtained in \cite{bob1}.

\subsection{The fermionic fields} 

   The six-dimensional $N=4$ gauged theory \cite{romans6} has
spin-$\ft12$ fields $\hat\chi$ and spin-$\ft32$ fields $\hat\psi_{M}$,
where the fermions carry also $USp(2)\times USp(2)$ indices, which we
are suppressing.  Thus we can think of the fermions as being tensor
products of 8-component spacetime spinors with two-component
$USp(2)\sim SU(2)$ spinors.  Now, we shall denote the six-dimensional
spacetime Dirac matrices by $\hat\gamma_A$.  It is necessary to
distinguish these hatted $8\times 8$ matrices from the $4\times 4$
Dirac matrices of the reduced theory in $D=5$, which will be denotes
by $\gamma_a$ without hats.  We may take a basis where the spacetime
Dirac matrices are related by
\be
\hat \gamma_a= \sigma_1\times\gamma_a\,,\qquad \hat\gamma_6=
\sigma_2\times \oneone\,,
\ee
where $\sigma_1$ and $\sigma_2$ are Pauli matrices.  Note that the
chirality operator in $D=6$ is given in this basis by
\be
\hat\gamma_7 = \sigma_3\times\oneone\,.
\ee

   In the
bosonic backgrounds that we need to consider, where the dilaton
$\hat\phi$ vanishes, the six-dimensional supersymmetry transformation
rule for the spin-$\ft12$ fields is given by
\be
\delta\hat\chi = -\ft{\im}{24}\, \hat F^{MNP}\, 
\hat\gamma_7\, \hat\gamma_{MNP}\,\hat\ep  - \ft1{8\sqrt2}\, 
\hat\gamma^{MN}\, (\hat G_{MN} +\im\,  \hat F^i_{MN}\, \hat\gamma_7\, 
\tau_i)\, \hat\ep\,,\label{d6chi}
\ee
where $\tau_i$ are the Pauli matrices associated with the internal
$USp(2)$ 2-component index.
The $D=6$ gravitino transformation rule, after setting $\hat\phi$ to
zero, is
\bea
\delta\, \hat\psi_M &=& \hat D_M\, \hat\ep - \ft{\im}{2}\,k\, 
\hat\gamma_M\, \hat\gamma_7\, \hat\ep - \ft1{48} \hat\gamma_7\,
\hat F_{NPQ}\, \hat\gamma^{NPQ}\,\hat\gamma_M\, \hat \ep\nn\\
&& -\ft{\im}{16\sqrt2}
(\hat\gamma_M{}^{PQ} - 6 \delta_M^P\, \hat\gamma^Q)\, (\hat G_{PQ}
+\im\, \hat F_{PQ}^i\, \hat\gamma_7\, \tau_i)\, \hat\ep\,,\label{d6psi}
\eea
where
\be
\hat D_M\, \hat\ep = \hat\nabla_M\,\hat\ep - \ft{\im}2 g\, \hat A_M^i\,
\tau_i\, \hat\ep\,.
\ee
In our case where only one component is of the Yang-Mills fields $\hat
F_\2^i$ is non-zero, we shall take it to be $i=3$.

    Our fermionic ansatz involves setting $\hat\chi=0$, and so
supersymmetry requires $\delta\hat\chi=0$, and hence, after substituting
(\ref{d6ans}) into (\ref{d6chi}), we get
\be
\Big(\im\, \cosh(k\, z)\, \hat\gamma_7\, \hat\gamma_6 +\sinh(k\, z) -
\im\, \hat\gamma_7\, \tau_3\Big)\, \hat\ep=0\,.\label{d6chicon}
\ee
Our reduction ansatz also involves setting the $z$ component of the
six-dimensional gravitino to zero.  From the $z$ component of the
gravitino transformation rule (\ref{d6psi}), we find then that for
surviving $D=5$ supersymmetry $\hat\ep$ should satisfy
\be
\del_z\, \hat\ep - \ft{\im}{2}\, k\, \hat\gamma_6\, \hat\gamma_7\,
\hat\ep=0\,.
\ee
 From these equations, we find that the Kaluza-Klein ansatz for
$\hat\ep$ should be
\be
\hat\ep\, =\Big( \cosh(\ft12 k\, z)\ -\im\, \sinh(\ft12k\, z)\, \hat\gamma_7\,
\hat\gamma_6\Big) \,\ep\,,
\ee
where $\ep$ is the five-dimensional supersymmetry parameter, which
must satisfy the projection condition
\be
- \tau_3\, \ep = \sigma_2\, \ep\,.
\ee
This condition halves the number of components of supersymmetry in
$D=5$, as we should expect since we are ending up with $N=2$ gauged
supergravity. 

    We find that the Kaluza-Klein reduction ansatz for $\hat\psi_\mu$
is
\be
\hat\psi_\mu\, =\Big(\cosh(\ft12 k\, z)\ -\im\, \sinh(\ft12k\, z)\, 
\hat\gamma_7\, \hat\gamma_6\Big) \,\psi_\mu\,,
\ee
where $\psi_\mu$ is the five-dimensional gravitino, which must also
satisfy the projection condition
\be
-\tau_3\, \psi_\mu =  \sigma_2\, \psi_\mu\,.
\ee
Substituting into the previous equations, this gives rise to the
following five-dimensional gravitino transformation rule:
\be
\delta\,\psi_\mu = D_\mu\, \ep - \ft{1}{2}\, k\, \sigma_2\, \gamma_\mu
\, \ep - \ft{\im}{8\sqrt3}\, F_{\nu\rho}\, (\gamma_\mu{}^{\nu\rho} - 4
\delta_\mu^\nu\, \gamma^\rho)\, \ep\,,
\ee
where
\be
D_\mu\, \ep \equiv \nabla_\mu\, \ep +\ft{\im}2 g\, \sqrt{\ft23}\, A_\mu\,
\tau_3\, \ep\,.
\ee

\subsection{Lifting to massive type IIA in $D=10$}

   The ansatz for obtaining the
bosonic sector of six-dimensional $SU(2)$ gauged $N=4$ supergravity 
as a consistent Kaluza-Klein reduction from massive type IIA
supergravity in $D=10$ was found in \cite{d6gauge}.  The Lagrangian
describing the bosonic sector of the massive IIA theory is
\bea
{\cal L}_{10} &=& \hat R\, {\hat *\oneone} -
\ft12 {\hat *d\hat \phi}\wedge d\hat \phi
- \ft12 e^{\fft32\hat\phi}\, {\hat *\hat F_\2}\wedge \hat F_\2 -
\ft12 e^{-\hat \phi}\, {\hat *\hat F_\3}\wedge \hat F_\3-
\ft12 e^{\fft12\hat\phi}\, {\hat *\hat F_\4}\wedge \hat F_\4 \nn\\
&&\!\!
-\ft12 d\hat A_\3\wedge d\hat A_\3 \wedge \hat A_\2 - \ft16 m\,
d\hat A_\3 \wedge (\hat A_\2)^3
 -\ft1{40} m^2\, (\hat A_\2)^5 -\ft12 m^2\, e^{\fft52\hat \phi}\,
{\hat *\oneone}\,,\label{romans1}
\eea
where the field strengths are given in terms of potentials by
\bea
\hat F_\2 &=& d\hat A_\1 + m\, \hat A_\2\ ,\qquad \hat F_\3 =
d\hat A_\2\,,\nn\\
\hat F_\4 &=& d\hat A_\3 + \hat A_\1\wedge d\hat A_\2 + \ft12 m\,
\hat A_\2\wedge \hat A_\2\,,\label{romfields}
\eea
and in this subsection we are using the hat symbol to denote the
ten-dimensional fields and Hodge dual.  

   It was shown in \cite{d6gauge}
that the consistent Kaluza-Klein reduction ansatz is
\bea
d\hat s_{10}^2\!\! &=&\!\! s^{\fft1{12}}\, X^{\fft18}\Big[
\Delta^{\fft38}\, ds_6^2 + 2g^{-2}\, \Delta^{\fft38}\, X^2\, d\xi^2
+\ft12g^{-2}\, \Delta^{-\fft58}\, X^{-1}\, c^2\, 
\sum_{i=1}^3(\sigma^i - g\, A_\1^i)^2\Big]\,,\nn\\
\hat F_\4 &=& -\ft{\sqrt2}{6}\, g^{-3}\, s^{1/3}\, c^3\, \Delta^{-2}\,
U\, d\xi\wedge\ep_\3 -\sqrt2 g^{-3}\, s^{4/3}\, c^4\, \Delta^{-2}\,
X^{-3}\, dX\wedge \ep_\3 \nn\\
&&-\sqrt2 g^{-1}\,
s^{1/3}\, c\, X^4\, {*F_\3}\wedge d\xi
-\ft1{\sqrt2} s^{4/3}\, X^{-2}\, {*G_\2} \nn\\
&& +\ft1{\sqrt2} g^{-2}\,
s^{1/3}\, c\, F_\2^i \, h^i\wedge d\xi -\ft1{4\sqrt2} g^{-2}\,
s^{4/3}\, c^2\, \Delta^{-1}\, X^{-3}\,  F_\2^i \wedge
h^j\wedge h^k\, \ep_{ijk}\,,\label{fans}\\
\hat F_\3 &=& s^{2/3}\, F_\3 + g^{-1}\, s^{-1/3}\, c\, G_\2\wedge d\xi
\,,\nn\\
\hat F_\2 &=& \ft1{\sqrt2}\, s^{2/3}\, G_\2\,,\qquad
e^{\hat\phi} = s^{-5/6}\, \Delta^{1/4}\, X^{-5/4}\,,\nn
\eea
where $X$ is related to the
six-dimensional dilaton $\phi$ by $X=e^{-\fft1{2\sqrt2}\phi}$, and
\be
\Delta \equiv  X\, c^2 +X^{-3} s^2 \,,\quad
U \equiv X^{-6}\, s^2 - 3 X^2\, c^2 + 4 X^{-2}\, c^2 - 6 X^{-2}\,.
\ee
We also define $h^i\equiv \sigma^i-g\, A_\1^i$, $\ep_\3\equiv
h^1\wedge h^2\wedge h^3$, and $s=\sin\xi$ and $c=\cos\xi$.  The
unhatted quantities, and Hodge dual $*$, refer to the six-dimensional
fields.  These fields satisfy the equations of motion following from
(\ref{d6lag}) (with the hats dropped on all the quantities in
(\ref{d6lag})), by virtue of the ten-dimensional equations of motion
following from (\ref{romans1}).  

   Substituting our braneworld reduction ansatz (\ref{d6ans}) into 
(\ref{fans}), we therefore find that the following gives a consistent 
braneworld Kaluza-Klein reduction from massive type IIA supergravity
in $D=10$ to $N=2$ gauged supergravity in $D=5$:
\bea
d\hat s_{10}^2\!\! &=&\!\! s^{\fft1{12}}\, \Big[
dz^2 + \cosh^2(k\, z)\, ds_5^2 + 2g^{-2}\, d\xi^2
+\ft12g^{-2}\, c^2\, (\sigma_1^2 + \sigma_2^2 + 
(\sigma_3 +\sqrt{\ft23}\,  g\, A_\1)^2)\Big]\,,\nn\\
\hat F_\4\!\!\! \!&=& \!\!\!\! -\ft{\sqrt2}{6}\, g^{-3}\, s^{1/3}\, c^3\,
d\xi\wedge\sigma_1\wedge\sigma_2\wedge (\sigma_3+\sqrt{\ft23}\, g\, A_\1) 
 +\sqrt{\ft23}\,  g^{-1}\,
s^{1/3}\, c\, \cosh^2(k\, z)\,  {*F_\2}\wedge d\xi\nn\\
&&+\ft1{\sqrt3} s^{4/3}\, \sinh(k\, z)\, \cosh(k\, z)\,  {*F_\2}\wedge
dz -\ft1{\sqrt3} g^{-2}\,
s^{1/3}\, c\, F_\2 \wedge(\sigma_3+\sqrt{\ft23}\, g\, A_\1)\wedge d\xi\nn\\
&& +\ft1{2\sqrt3} g^{-2}\,
s^{4/3}\, c^2\,  F_\2 \wedge \sigma_1\wedge\sigma_2\,,\label{105ans}\\
\hat F_\3 &=& -\ft1{\sqrt3}\, s^{2/3}\, \cosh(k\, z)\, dz\wedge F_\2  
 -\sqrt{\ft23} \, g^{-1}\, s^{-1/3}\, c\, \sinh(k\, z)\, F_\2\wedge d\xi
\,,\nn\\
\hat F_\2 &=& -\ft1{\sqrt3}\, s^{2/3}\, \sinh(k\, z)\, F_\2\,,\qquad
e^{\hat\phi} = s^{-5/6}\,,\nn
\eea
where we recall that $k=g\, \sqrt2/3$.

\section{Braneworld reductions from type IIB$^*$ and IIA$^*$}

    Hull has proposed that the theories one obtains by performing a
T-duality transformation of type IIA or type IIB supergravity with a 
timelike reduction should be viewed as low-energy limits of 
consistent sectors of string theory \cite{hull1}.  These theories,
which he calls type IIB$^*$ and type IIA$^*$ respectively, differ from 
the usual type IIB and IIA theories in that the signs of the kinetic
terms of all the Ramond-Ramond fields are reversed.  In fact, the
type IIB$^*$ and type IIA$^*$ theories can be obtained from the usual type
IIB and type IIA theories by making the replacements
\be
\Psi\longrightarrow \im\, \Psi\,,\label{tostar}
\ee
where $\Psi$ denotes the set of all Ramond-Ramond fields.  Recently,
gauged de Sitter supergravities that are related to the 
type IIA$^*$ and type IIB$^*$ theories were discussed \cite{hull2}.

   In the rest of this section, we shall investigate the possibility
of performing braneworld Kaluza-Klein reductions based on these type
IIB$^*$ and IIA$^*$ theories.  Since the theories themselves can be
obtained by the analytic continuation (\ref{tostar}), we can obtain
the associated braneworld reductions by performing appropriate
analytic continuations of our previous results.  As we shall see, we
can, by this means, obtain a braneworld reduction giving
four-dimensional $N=2$ gauged de Sitter supergravity on the brane.  We
also investigate the analogous procedure for a braneworld reduction to
five-dimensional $N=2$ gauged de Sitter supergravity, and show that in
this case we cannot obtain a theory with real fields.

\subsection{$N=2$, $D=4$ de Sitter supergravity on the brane}
\label{d4desitter}

      We first consider the braneworld reduction of gauged $N=4$, $D=5$ 
de Sitter supergravity.  Specifically, this five-dimensional
supergravity will itself be obtained as a reduction from $D=10$, but
now arising as a reduction of type IIB$^*$ supergravity on $H^5$, the
hyperbolic 5-space.  Implementing (\ref{tostar}), we need to make the
replacements
\be
\hat H_\5\longrightarrow \im\, \hat H_\5\,,\qquad
\hat A_\2^1 \longrightarrow \im\, \hat A_\2^1\,,\qquad
\hat\chi\longrightarrow \im\, \hat\chi
\ee
in the type IIB Lagrangian (\ref{d10lag}).  Since our goal is to get a
reduction to $D=5$ after having made these replacements, it suffices
for us to implement this directly in the reduction ansatz
(\ref{d105ans}).  We can achieve this with the following replacements:
\bea
&&g\longrightarrow \im\, g\,,\qquad \xi\longrightarrow \im\, \xi +
\ft12\pi\,,\qquad \tau\longrightarrow \im\, \tau + \ft12 \pi\,,\nn\\
&&A_\2^1 \longrightarrow \im\, A_\2^1\,,\qquad
A_\1^i\longrightarrow \im\, A_\1^i\,.
\eea
It is easily seen that this gives a real Kaluza-Klein reduction ansatz
under which the type IIB$^*$ theory reduces on $H^5$, to yield a de
Sitter supergravity in $D=5$ (see also \cite{hull2}).  In particular 
the metric reduction ansatz in (\ref{d105ans}) has become
\bea
d\hat s^2_{10} &=&
\Delta^{1/2}\, ds_5^2 +  g^{-2} \, X\,
\Delta^{1/2}\, d\xi^2 +
g^{-2}\Delta^{-1/2}\, X^2\, \td c^2\, \Big(d\tau - g\, B_\1\Big)^2 \nn\\
& &+ \ft14 g^{-2}\, \Delta^{-1/2}\, X^{-1}\,
\td s^2\, \sum_i (\sigma^i + \sqrt{2}g\, A_\1^i)^2\,,
\eea
where $\td s\equiv \sinh\xi$, $\td c\equiv \cosh\xi$, and $\Delta\equiv
X^{-2}\, \td c^2-X\, \td s^2$. Note that the level surfaces at constant 
$\xi$ in the internal dimensions are now $\R\times S^3$ (as opposed to
$S^1\times S^3$ before the analytic continuation), giving rise to an
$H^5$ topology (as opposed to the $S^5$ topology before the analytic
continuation).  The five-dimensional theory is gauged $N=4$ de Sitter
supergravity, with $\R\times SU(2)$ Yang-Mills fields.  We have seen 
that it is obtained by making the replacements
\be
g\longrightarrow \im\, g\,, \qquad \hat A_\2^1\longrightarrow \im\,
\hat A_\2^1\,,\qquad \hat A_\1^i\longrightarrow \im\, \hat A_\1^i
\label{d5ds}
\ee
in the five-dimensional Lagrangian (\ref{d5lag}).  

    The next step is to perform the appropriate analytic continuation 
of the braneworld reduction (\ref{d5ansatz}), in order to obtain a
reduction of the five-dimensional theory coming from (\ref{d5lag}) 
with the replacements (\ref{d5ds}). It is easily seen that {\it in
terms of the analytically continued quantities} defined in
(\ref{d5ds}) for the
five-dimensional de Sitter supergravity, the braneworld reduction
ansatz (\ref{d5ansatz}) will become
\bea
d\hat s_5^2 &=& dz^2 + \cos^2(k\, z)\, ds_4^2\,,\nn\\
\hat A_\2^1 &=& -\ft1{\sqrt2}\,  \cos(k\, z)\, {*F_\2}\,,\qquad 
\hat A_\2^2 = -\ft1{\sqrt2}\, \sin(k\, z)\, F_\2\,,\nn\\
\hat A_\1^1 &=& \ft1{\sqrt2}\, A_\1\,,\label{d5dsansatz}
\eea
where we also have
\be
k=g\,.
\ee
The equations of motion of the five-dimensional de Sitter supergravity
imply the following four-dimensional equations
\bea
R_{\mu\nu} &=& -\ft12 (F_{\mu\rho}\, F_\nu{}^\rho - 
         \ft14 F_\2^2 \, g_{\mu\nu}) + 3k^2\, g_{\mu\nu}\,,\nn\\
d{*F_\2} &=& 0\,,\label{d4dseom}
\eea
which are the bosonic equations of motion for gauged $N=2$ {\it de
Sitter} supergravity in $D=4$.  Note that in comparison to the
previous AdS braneworld reduction ansatz (\ref{d5ansatz}), the
four-dimensional Maxwell potential $A_\1$ has itself undergone an
analytic continuation $A_\1\longrightarrow \im\, A_\1$.  This has the
consequence, apparent in the Einstein equation in (\ref{d4dseom}),
that it has the ``wrong sign'' for its kinetic term.  This is indeed
one of the features of $N=2$ gauged de Sitter supergravity in $D=4$.
It should also be remarked that the cosmological constant in $D=4$ can
again be adjusted freely, whilst holding the five-dimensional
cosmological constant fixed, by making rescalings as in (\ref{rescale}).
It is straighforward also to obtain the reduction ansatz for the 
fermionic fields, by implementing the analytic continuations on the
reductions obtained in section \ref{d54fermion}.  We shall discuss
the fermions further in section \ref{ds4ads5}.

\subsection{$N=2$, $D=5$ de Sitter supergravity on the brane?}
\label{ds6ds5}

    We can now repeat the analogous steps for the reduction
massive type IIA$^*$ supergravity to give $N=4$ gauged de Sitter
supergravity in $D=6$, and then investigate the possibility of 
a braneworld reduction to $N=2$ gauged de Sitter supergravity in
$D=5$.  The first part of this procedure is a straightforward
generalisation of the one followed in the previous subsection.  We
make the following analytic continuation in the massive type IIA
reduction ansatz (\ref{fans}),
\bea
&&g\longrightarrow \im\, g\,,\qquad
\xi\longrightarrow \ft12\pi + \im\, \xi\, \nn\\
&&B_\1\longrightarrow \im\, B_\1\,,\qquad A_\1^i\longrightarrow 
\im\, A_\1^i\,,
\eea
which leads to a consistent reduction to $N=4$ gauged de Sitter
supergravity in $D=6$.  Note that we still get a real theory after the
reduction, even though fractional powers of
$\sin\xi$ appear in the reduction ansatz, since it is only the
$\cos\xi$ terms that pick up factors of $\im$, and these are all
raised to integer powers. 
The resulting six-dimensional de Sitter supergravity is therefore described by
the Lagrangian (\ref{d6lag}), after making the analytic continuations
\be
g\longrightarrow \im\, g\,,\qquad
\hat B_\1\longrightarrow \im\, \hat B_\1\,,\qquad \hat A_\1^i\longrightarrow 
\im\, \hat A_\1^i\,.\label{d5dsrep}
\ee

   However, when we proceed to the next stage, of looking for an
analytic continuation of the previous braneworld reduction ansatz
(\ref{d6ans}), we encounter a difficulty.  This can be attributed to
the fact that compatibility between the $SU(2)$ Yang-Mills ansatz in
(\ref{d6ans}) and (\ref{d5dsrep}) will require that we must make the
continuation $A_\1\longrightarrow \im\, A_\1$ in the 5-dimensional
Maxwell potential.  However, this conflicts with the reality
conditions for the remainder of the reduction ansatz.  In fact the
reason for this can be seen by looking at the equation of motion for
the Maxwell field in (\ref{d5sugra}); the non-linear term on the
right-hand side means that we cannot perform the analytic continuation 
$A_\1\longrightarrow \im\, A_\1$ and still get a real five-dimensional
theory.  This is quite different from the situation in four
dimensions, where there is no analogous Chern-Simons term preventing
one from sending $A_\1$ to $\im\, A_\1$.   

   If we wanted to get a real de Sitter theory in $D=5$ via a braneworld
reduction from $D=6$, we could consider the following $D=6$ to $D=5$
ansatz,
\bea
&&d\hat s_6^2 = dz^2 + \cos^2(k\, z)\, ds_5^2\,,\nn\\
&&\hat A_\2 = -\ft1{\sqrt3}\, k^{-1}\, \sin(k\, z)\, F_\2\,,\nn\\
&&\hat A_\1^3 =\im\, \sqrt{\ft23}\, A_\1\,,\nn\\
&&\hat B_\1 =0\,,\qquad \hat A_\1^1=0\,,\qquad \hat A_\1^2=0\,,\qquad 
\hat\phi=0\,,
\eea
with $k=\sqrt2\, g/3$. This does formally provide a reduction to a
real $D=5$ de Sitter theory, although with the price that the ansatz in
$D=6$ is complex.  Thus the situation is quite different from that in
section \ref{d4desitter}, where we obtained a a completely real
reduction to get $N=2$ gauged de Sitter supergravity on the
four-dimensional brane.

\section{de Sitter supergravity on the brane from anti-de Sitter}
\label{ds4ads5}

   It is also possible to embed the $D$-dimensional de Sitter metric
in the $(D+1)$-dimensional anti-de Sitter metric, rather than in the
de Sitter metric, by replacing $\cosh(k\, z)$ by $\sinh(k\, z)$ in
(\ref{adsdd1});
\be
d\hat s_{D+1}^2 = dz^2 + \sinh^2(k\, z)\, ds_D^2\,.
\ee
This was discussed in the context of four-dimensional braneworld
models in \cite{karran}.

   We find that we can implement this as a consistent braneworld
Kaluza-Klein reduction, starting from and $N=4$ AdS gauged
supergravity in five dimensions in which the two 2-form potentials are
analytically continued by multiplication by $\im$, to obtain the same
$N=2$ gauged de Sitter supergravity in $D=4$ that we discussed in
section \ref{d4desitter}.  After first replacing the potentials 
$\hat A_\2^\a$ in the usual $N=4$ gauged supergravity in $D=5$ by 
$\im\, \hat A_\2^\a$, we obtain another real theory, for which
the bosonic braneworld reduction ansatz is given by
\bea
d\hat s_5^2 &=& dz^2 + \sinh^2(k\, z)\, ds_4^2\,,\nn\\
\hat A_\2^1 &=& -\ft1{\sqrt2}\, \sinh(k\, z)\, {*F_\2}\,,\qquad 
\hat A_\2^2 = \ft{1}{\sqrt2}\, \cosh(k\, z)\, F_\2\,,\nn\\
\hat A_\1^1 &=& \ft1{\sqrt2}\, A_\1\,,\label{d5ansatzx}
\eea
and this consistently yields the de Sitter gauged supergravity
equations (\ref{d4dseom}).  

    The above result can in fact be obtained directly from our
previous anti-de Sitter braneworld reduction in section \ref{bossec}.
This is done by making the coordinate transformation $z\longrightarrow
z + \im\, \pi/(2k)$, together with sending $ds_4^2\longrightarrow -
ds_4^2$, in the expressions in section \ref{d4n2sugra}.  This can
indeed be seen to give a real embedding, provided that the 2-form
potentials in five dimensions are analytically continued as described
above, namely $\hat A_\2^\a\longrightarrow \im\, \hat A_\2^\a$.  

   This five-dimensional AdS gauged supergravity can itself be
obtained by dimensional reduction from a variant of type IIB
supergravity.  Specifically, we take as our starting point the type
IIB$_{7+3}$ theory introduced in \cite{hull3}.  This has spacetime
signature $(7,3)$, and it arises from a sequence of T-duality
transformations of the usual type IIB theory, using timelike as well
as spacelike circles in the various transformation steps.  It has
symplectic Majorana-Weyl spinors, and the kinetic terms for the two
2-form potentials are of the non-standard sign, whilst all other
bosonic fields have standard-sign kinetic terms.  It can be reduced on
the maximally-symmetric space $O(4,2)/O(3,2)$ of signature $(3,2)$, to
give the non-standard AdS gauged supergravity in $D=5$ that we invoked
in our braneworld construction above.  The reduction of the type
IIB$_{7+3}$ theory on $O(4,2)/O(3,2)$ can be derived from the $S^5$
reduction of type IIB in (\ref{d105ans}), by first analytically
continuing the 2-form potentials $\hat A_\2^\a\longrightarrow \im\,
\hat A_\2^\a$, and then sending $\xi\longrightarrow \im\, \xi$.  
It is straightforward to extend the discussion to include the
fermions too, to get the complete four-dimensional de Sitter gauged
$N=2$ supergravity.  

   One can also consider the possibility of a similar embedding 
of five-dimensional de Sitter gauged supergravity in six-dimensional 
anti-de Sitter gauged supergravity.  However, in this case we can see
from (\ref{d6ans}) that after sending $z\longrightarrow z + \im \,
\pi/(2k)$ we would need also to send $A_\1\longrightarrow \im A_\1$ in
order to keep $\hat A_\2$ real, but this would then contradict the
ansatz for the Yang-Mills potential $\hat A_\1^3$, which requires that
$A_\1$ be kept untransformed.  Thus we see in this approach, just like
the de Sitter to de Sitter reduction considered in section
\ref{ds6ds5}, we cannot get a real embedding of five-dimensional de
Sitter gauged supergravity on the brane.

\section{Conclusions}

    In this paper, we have shown how the results in \cite{bob1,bob2}
on the braneworld Kaluza-Klein reductions to ungauged supergravities
can be extended, in certain cases, to braneworld reductions giving
gauged supergravities.  Specifically, we have constructed such
reductions from five-dimensional $N=4$ gauged supergravity to
four-dimensional $N=2$ gauged supergravity on the brane, and likewise
from six-dimensional $N=4$ gauged supergravity to five-dimensional
$N=2$ gauged supergravity on the brane.  The lower-dimensional
cosmological constant is freely adjustable, whilst holding the
higher-dimensional one fixed, and so one can, for example,  
arrange to have a phenomenologically realistic cosmological constant 
in four dimensions ``on the brane,'' even if the five-dimensional
cosmological constant is of the Planck scale. In each case, the
higher-dimensional starting point can itself be obtained as a
consistent Kaluza-Klein sphere reduction, from type IIB supergravity
in the first example \cite{d5gauge}, and from massive type IIA in the
second \cite{d6gauge}.  Thus one can lift the braneworld embeddings to
the corresponding ten-dimensional theories.

    The braneworld reductions allow one to obtain new explicit exact
solutions of higher-dimensional supergravities, by starting from
known solutions of the gauged supergravities on the brane.  Thus, for
example, we can consider a charged AdS black hole solution of the
$N=2$ gauged supergravity in $D=4$, and lift it to a solution of the
$N=4$ gauged supergravity in $D=5$.  The charged AdS$_4$ black hole is
given by \cite{dufliu,sabra}
\bea
&&ds_4^2 = -H^{-2}\, f\, dt^2 + H^2\, (f^{-1}\, dr^2 + r^2\,
d\Omega_{2,\ep}^2)\,,\label{ads4bh}\\
&&A_\1= \ft12\sqrt{\ep}\, (1-H^{-1})\, \coth\beta\, dt\,,\quad
H=1 + \fft{\mu\, \sinh^2\beta}{\ep\, r}\,,\quad 
f= \ep-\fft{\mu}{r} + 4g^2\, r^2\, H^4\,,\nn
\eea
where $\mu$ and $\beta$ are constants, and $\ep=1$,0 or -1 according
to whether the foliations in the space transverse to the black hole
have the metric $d\Omega_{2,\ep}^2$ on the unit $S^2$, $T^2$ or
hyperbolic space $H^2$.  (In the case $\ep=0$ one must scale
$\sinh^2\beta\longrightarrow \ep\, \sinh^2\beta$ before sending $\ep$
to zero.)  (The AdS$_4$ black hole is written here in the notation of
\cite{ten}, and corresponds to setting all four charges equal in
the four-charge solution given there.) Substituting the solution
(\ref{ads4bh}) into (\ref{d5ansatz}) gives an embedding of the AdS$_4$
black hole in five-dimensional $N=4$ gauged supergravity, whilst
substituting it into (\ref{d104ans}) gives an embedding of the AdS$_4$
black hole in type IIB supergravity.  One can similarly embed an
AdS$_5$ black hole \cite{becvsa} in six-dimensional $N=4$ gauged
supergravity using (\ref{d6ans}), and then in massive type IIA using
(\ref{105ans}).  

   We then showed that, based on the embedding dS$_D\subset$ dS$_{D+1}$,
the braneworld reduction to four-dimensional $N=2$ gauged (anti-de
Sitter) supergravity could be analytically continued to give a
braneworld reduction to $N=2$ de Sitter gauged supergravity.  This
theory, and its five-dimensional de Sitter supergravity progenitor,
differ from normal gauged supergravities in having non-standard signs
for the kinetic terms of certain gauge fields.  (This is how they
manage to evade the usual theorems about the non-existence of de
Sitter supergravities.)  We showed that the five-dimensional de Sitter
supergravity in question can itself be obtained as a consistent
reduction (on the hyperbolic space $H^5$) of the type IIB$^*$
supergravity discussed by Hull \cite{hull1}.  This theory has the
non-standard sign for the kinetic terms of all the Ramond-Ramond
fields, and was obtained by performing a timelike T-duality
transformation of the type IIA theory \cite{hull1}.  It can
equivalently be thought of as an analytic continuation of the usual
type IIB theory in which all the Ramond-Ramond fields are multiplied
by a factor of $\im$.

   A similar continuation yields a consistent reduction of massive
type IIA$^*$ supergravity to give a de Sitter gauged supergravity in
six dimensions. In this case, by contrast, we do not get a real embedding
to give a five-dimensional $N=2$ braneworld de Sitter supergravity.

   We also showed that a very different braneworld Kaluza-Klein
reduction to give four-dimensional de Sitter gauged $N=2$ supergravity
can be achieved, in which we start from five-dimensional anti-de
Sitter gauged $N=4$ supergravity, with an analytic continuation of the
two 2-form potentials.  This depends upon the fact that one can embed
dS$_D$ in AdS$_{D+1}$.  The five-dimensional gauged theory can itself 
be derived from the type IIB$_{7+3}$ theory introduced in \cite{hull3}.
It is interesting that we can
obtain the four-dimensional de Sitter gauged $N=2$ supergravity by two
different routes, one of which has its ultimate origin in the
ten-dimensional type IIB$^*$ theory, and the other in the type
IIB$_{7+3}$ theory.

   In this paper we have obtained consistent braneworld reductions
based on each of the embeddings AdS$_D\subset$ AdS$_{D+1}$,
dS$_D\subset$ dS$_{D+1}$ and dS$_D\subset$ AdS$_{D+1}$.  It is worth
remarking that there is no fourth possibility, of having an embedding
AdS$_D\subset$ dS$_{D+1}$.  This can be seen easily from the fact that
the isometry groups of AdS$_n$ and dS$_n$ are $SO(n-1,2)$ and $SO(n,1)$ 
respectively, and that $SO(D-1,2)$ is not a subgroup of $SO(D,1)$.  

   Several interesting questions about the braneworld Kaluza-Klein
reductions we have considered in this paper arise.  For example, we
have seen that we obtain consistent reduction ans\"atze that give us
the massless gauged supergravities ``on the brane'' in the lower
dimension.  In particular, these supergravities include, of course,
the massless graviton in their spectrum of states.  This appears at
first sight to be at odds with the findings in \cite{karran} (see also
\cite{porrati}), where it is shown that there is no massless graviton
in the effective spectrum of states on the brane.  In fact it is not
clear precisely what the relation between the two viewpoints is.  This
question already arose in the braneworld reductions to ungauged
supergravity, constructed in \cite{bob1,bob2}, although in a slightly
less extreme form there since the existence of a massless graviton is
a feature common to both viewpoints in those examples.

   In the ungauged reductions the discrepancies between the standard
Randall-Sundrum viewpoint and the braneworld reduction viewpoint show
up, for example, in the discussion of the global structure of
solutions ``on the brane.''  Thus, for instance, in \cite{hawcha} a
four-dimensional Schwarzschild black hole was shown to give rise to a
solution in $D=5$ that was singular on the horizon of the AdS$_5$.
The absence of supersymmetry in that example provided a possible
mechanism, via a Gregory-Laflamme instability, for mitigating the
effects of this singularity.  However, in the braneworld reductions in
\cite{bob1,bob2}, the localised supergravities admit supersymmetric
black hole solutions, which presumably are not subject to such an
instability, and so understanding the singularities at the
AdS$_5$ horizon becomes more pressing.\footnote{One does not need a
{\it singular} or high-curvature solution on the brane in order to get
such a curvature singularity on the AdS$_5$ horizon; even a mild
gravitational disturbance on the brane will get ``amplified'' to give
a singularity far from the brane.}  A mechanism, within the
Randall-Sundrum framework, for eliminating such singularities was
proposed in \cite{gidkatran}.  It was argued that in practice the
distribution of energy from a disturbance on the brane would be
predominantly in very low mass, rather than massless, gravitons, and
the effects of these would decay with distance from the brane.
However, the existence of the exact Kaluza-Klein reductions obtained
in \cite{bob1,bob2} still seems to require explanation, since these
embeddings show that one can have arbitrarily large excitations of the
massless fields on the brane that will never (classically, at least)
``spill over'' into the light massive modes.

   It is interesting to note that the ``amplification'' effect that
causes the diverging curvature in the bulk in the ungauged embeddings
does not in fact occur in the gauged AdS embeddings, although it does
for gauged de Sitter embeddings.  To see this, we note that for a
Kaluza-Klein metric reduction of the form $d\hat s^2 = dz^2 + f^2\,
ds^2$, where $f$ depends on $z$, the curvature 2-forms are given by
\be
\hat\Theta_{0a} = -\fft{f''}{f}\, \hat e^0\wedge \hat e^a\,,\qquad 
\hat\Theta_{ab} = \Theta_{ab} -\fft{{f'}^2}{f^2}\, \hat e^a\wedge 
\hat e^b\,,
\ee
where $\hat e^0=dz$, $\hat e^a=f\, e^a$, $ds^2= e^a\otimes e^a$ and
$\Theta_{ab}$ are the curvature 2-forms for the lower-dimensional
metric $ds^2$.  Thus for the ungauged, gauged anti-de Sitter and
gauged de Sitter cases, the tangent-space components of the
higher-dimensional Riemann tensor $\hat R_{ABCD}$ are given by
\bea
\underline{\hbox{Ungauged}}:&& f=e^{-k\, z}\,,\qquad \hat R_{0a0b}=-k^2\,
\eta_{ab}\,,\nn\\
&&\hat R_{abcd}= e^{2k\, z} \, R_{abcd} -
k^2\, (\eta_{ac}\, \eta_{bd} -\eta_{ad}\, \eta_{bd})\,,\nn\\
&&\nn\\
\underline{\hbox{Gauged AdS}}:&& f=\cosh(k\, z)\,,\qquad \hat R_{0a0b}=-k^2\,
\eta_{ab}\,,\nn\\
&& 
  \hat R_{abcd}= \fft1{\cosh^2(k\, z)}\, R_{abcd} -
k^2\, \tanh^2(k\, z)\, 
  (\eta_{ac}\, \eta_{bd} -\eta_{ad}\, \eta_{bd})\,,\\
&&\nn\\
\underline{\hbox{Gauged dS (1)}}:&& f=\cos(k\, z)\,,\qquad \hat R_{0a0b}=k^2\,
\eta_{ab}\,,\nn\\
&&
 \hat R_{abcd}= \fft1{\cos^2(k\, z)}\, R_{abcd} -
k^2\, \tan^2(k\, z)\, 
  (\eta_{ac}\, \eta_{bd} -\eta_{ad}\, \eta_{bd})\,,\nn\\
&&\nn\\
\underline{\hbox{Gauged dS (2)}}:&& f=\sinh(k\, z)\,,
\qquad \hat R_{0a0b}=-k^2\, \eta_{ab}\,,\nn\\
&&
 \hat R_{abcd}= \fft1{\sinh^2(k\, z)}\, R_{abcd} -
k^2\, \coth^2(k\, z)\, 
  (\eta_{ac}\, \eta_{bd} -\eta_{ad}\, \eta_{bd})\,,\nn
\eea
where $R_{abcd}$ are the tangent-space components of the Riemann
tensor for the metric $ds^2$ in the lower dimension.  The two de Sitter
cases refer to the dS$_D\subset$ dS$_{D+1}$ and dS$_D\subset$ AdS$_{D+1}$
embeddings used in sections 4 and 5 respectively.  Riemann tensor
divergences in the higher-dimensional bulk resulting from
curvature (even finite) on the brane therefore occur if the
prefactor of $R_{abcd}$ diverges.  This occurs on the
higher-dimensional AdS horizon $z = +\infty$ in the
ungauged case \cite{bob1,bob2}, at $z=\pi/(2k)$ in the
first gauged de Sitter case, and at $z=0$ in the second 
gauged de Sitter case.  But in the gauged AdS case, the prefactor
$(\cosh(k\, z))^{-2}$ is always $\le1$, and no such divergence occurs.

   In our present case, however, where gauged supergravities are
obtained on the brane, we do have the remaining puzzle about the
existence of the massless graviton.  It may well be that a resolution
of the ungauged braneworld reduction puzzles described above would
also indicate the resolution of the massless graviton puzzle.  In both
cases, the differences between the usual effective-gravity discussions
of, for example, \cite{ransun2,gidkatran} and \cite{karran}, and the
Kaluza-Klein approach of \cite{bob1,bob2} and this paper, are
concerned with whether one considers the distribution of excitations
over massless and massive modes, or if, on the other hand, one
considers only an exact embedding of the massless modes alone.  It
seems that more investigation is needed in order to reconcile the
viewpoints.  In the meantime we present our results for their
intrinsic interest, since they provide new insights into the subject
of consistent Kaluza-Klein reductions, and, in particular, they
provide ways of constructing new exact higher-dimensional solutions
from known lower-dimensional ones.  For example, in a recent
application the ungauged braneworld reductions of \cite{bob1,bob2}
were used in order to construct exact multi-membrane solutions in
seven-dimensional gauged supergravity \cite{liuwen}.

\section*{Acknowledgments}  

    We should like to thank Mirjam Cveti\v c, Dan Freedman, Gary Gibbons, 
Chris Hull and Hong L\"u for valuable discussions.

\end{document}